\documentstyle[11pt,newpasp,twoside]{article} 
\def\edcomment#1{\iffalse\marginpar{\raggedright\sl#1\/}\else\relax\fi} 
\reversemarginpar 

\begin{document} 
\title{Analysing Multi-Color Observations of Young Star Clusters in Mergers}

\author{Peter Anders, Uta Fritze--v. Alvensleben} 
\affil{Universit\"ats-Sternwarte G\"ottingen, Geismarlandstrasse 11, \\ 37083 G\"ottingen, Germany} 
\author{Richard de Grijs} 
\affil{Institute of Astronomy, Madingley Road, Cambridge, CB3 0HA, UK}

\section{Introduction} 

We present a new evolutionary synthesis (ES) tool specifically developed for the analysis of multi-color data of young star clusters (YSCs) in interacting galaxies. Our ESO/ST-ECF ASTROVIRTEL\footnote{The support given by ASTROVIRTEL, a Project funded by the European Commission under FP5 Contract No. HPRI-CT-1999-00081, is acknowledged.} project provides an unprecedented database of UV--optical--NIR observations from the {\sl HST}/WFPC2, {\sl HST}/NICMOS and {\sl VLT} data archives. Comparison of these data with ES models for clusters of various metallicities which include gaseous emission as well as various degrees of dust extinction, allows one to independently determine metallicities, ages and extinction for individual clusters. These, in turn, are the basis to derive the mass functions of the YSC systems and to predict the future evolution of their luminosity functions and color distributions. Comparing YSC systems of various ages will, we anticipate, allow us to ``see cluster disruption processes at work''. 

\section{Model Description} 

We have further refined the G\"ottingen ES code by including the effect of gaseous emission lines and continuum emission. The emission lines contribute significantly to the integrated light of stellar populations younger than $\sim 10^8$ years (up to 60\%, depending on age, metallicity and passband) (cf. Anders et. al. 2002a in prep.). We have also included the effects of {\sl internal} extinction. 

The simultaneous determination of cluster age, metallicity, and extinction is done by least-squares optimisation of an appropriate grid of our ES models with respect to the observed colors. 

Due to the well-known age-metallicity (and age-extinction) degeneracy for optical (and to some extent also NIR) colors, the use of multi-passband observations, in at least four passbands, is essential to determine these three free parameters independently. 

\subsection{Results for Artificial Clusters}

We constructed five artificial clusters of solar metallicity, extinction E(B-V)=0.1, and ages of 8, 60, 200 Myr, and 1 and 10 Gyr. The appropriate colors for these artificial clusters were taken from our models and a typical observational error of 0.1 mag was assumed. We subsequently redetermined the cluster parameters and compared these with the original values. From this exercise we conclude that:\\
{\it (i)} The passband combinations required for the recovery of the basic cluster parameters can be optimized towards the expected mean parameter values.\\
{\it (ii)} In order to determine reliable ages, the use of UV-passbands is essential for all ages. Including a NIR passband improves the accuracy.\\
{\it (iii)} For metallicity determination, NIR colors are most important, while the UV offers valuable information for young systems (ages $\la$ 100 Myr).\\
{\it (iv)} The extinction is best determined using UV + optical colors.

In general, the accuracies and the best passband combination are highly dependent on the age of the cluster's stellar population (see Anders et. al. 2002b in prep.). The availability of observations in the whole range, from UV to NIR, offers the possibility to constrain all independent parameters (age, metallicity and internal extinction) most effectively.

\section{Application to the Young Star Clusters in NGC 1569}

The irregular dwarf galaxy NGC 1569 is commonly classified as a (post-)starburst galaxy with a metallicity of $\sim (0.2-0.5) Z_\odot$. High-resolution multipassband data are provided by the ESO/ST-ECF ASTROVIRTEL project ``The Evolution and Environmental Dependence of Star Cluster Luminosity Functions'' (PI R. de Grijs). In addition to the well-studied super star clusters in its main disk, they reveal a number of fainter objects resembling compact star clusters. Our cluster fitting algorithm provides evidence for continuous star formation starting at least 1 Gyr ago. However, we detect a recent burst of cluster formation starting some 60 Myr ago, with a further enhanced cluster formation rate during the last 16 Myr and a peak formation rate in the youngest age bin ($\la 8$ Myr). These results are consistent with previous results focusing on the overall starburst activity, and on the formation of the two super star clusters in particular. As seen in previous studies, the most recent, enhanced epoch of cluster formation coincides closely with the formation of morphologically peculiar features, such as H$\alpha$ filaments and ``superbubbles''. 

An accompanying H{\sc i} cloud is located some 5 kpc from NGC 1569, and connected to it by an H{\sc i} bridge. This suggests a tidal interaction as a possible cause for the starburst. We see hints for propagation of cluster formation over the last few dozen Myr, starting from the H{\sc i} bridge. The masses of the clusters cover the range of $(400-230.000) M_\odot$. 

The cluster metallicity distribution is broader than the range in ISM abundances found in previous surveys, which mainly covered the ISM. This may imply different chemical enrichment processes in the star clusters vs. the ISM, or selection effects caused by the different spatial coverage. The internal extinction is found to be low ($E(B-V) \la 0.25$, with the majority of the clusters having $E(B-V) \la 0.1$) (Anders et. al. 2002c in prep.).

\end{document}